\documentclass[aps,pra,9pt,twocolumn,showpacs,superscriptaddress]{revtex4-2}

\usepackage{amsmath}
\usepackage{latexsym}
\usepackage{amssymb}
\usepackage{bm}
\usepackage{graphics,epstopdf}
\usepackage{color}

\usepackage{newlfont}
\usepackage{amsfonts}
\usepackage{amsthm}
\usepackage{graphicx}
\usepackage{epsfig}
\usepackage[colorlinks=true,linkcolor=blue,citecolor=red,plainpages=false,pdfpagelabels]{hyperref}

\DeclareOldFontCommand{\rm}{\normalfont\rmfamily}{\mathrm}

\newcommand{\norm}[1]{\left\lVert#1\right\rVert}

\newcommand{\ket}[1]{|{#1}\rangle}

\usepackage{times}

\newcommand{\be}{\begin{equation}}
\newcommand{\ee}{\end{equation}}
\newcommand{\bc}{\begin{center}}
\newcommand{\ec}{\end{center}}
\newcommand{\bea}{\begin{eqnarray}}
\newcommand{\eea}{\end{eqnarray}}
\newcommand{\ba}{\begin{array}}
\newcommand{\ea}{\end{array}}

\begin{document}

\title{Quantum Speed Limits for Observables}

\author{Brij Mohan\(^{}\)}\email{brijhcu@gmail.com}
\affiliation{\(\)Harish-Chandra Research Institute,\\  A CI of Homi Bhabha National
Institute, Chhatnag Road, Jhunsi, Prayagraj  211019, Uttar Pradesh, India
}

\author{Arun Kumar Pati\(^{}\)}\email{akpati@hri.res.in}
\affiliation{\(\)Harish-Chandra Research Institute,\\  A CI of Homi Bhabha National
Institute, Chhatnag Road, Jhunsi, Prayagraj  211019, Uttar Pradesh, India
}

\begin{abstract}
In the Schr{\"o}dinger picture, the state of a quantum system evolves in time  and the quantum speed limit describes how fast the 
state of a quantum system evolves from an initial state to a final state. However, in the Heisenberg picture the observable evolves in time instead of the state vector. Therefore, it is natural to ask how fast an observable evolves in time. This can impose a fundamental bound on the evolution time of the expectation value of quantum mechanical observables. We obtain the quantum speed limit time-bound for observable 
for closed systems, open quantum systems and arbitrary dynamics. Furthermore, we discuss various applications of these bounds. Our results can have several applications ranging from setting the speed limit for operator growth, correlation growth, quantum thermal machines, quantum control and many body physics. 
\end{abstract}


\maketitle
\section{Introduction}
Time is one of the fundamental notions in the physical world, and it plays a significant role in almost every existing physical theory. Yet, understanding time has been a challenging task and often it is treated like parameter. Even though time is not an operator,  there is a geometric uncertainty relation between time and energy fluctuation which imposes inherent limitation on how fast a quantum system can evolve in time. This was first discovered in an attempt to operationalize the time-energy uncertainty relation. This concept is now known as the quantum speed limit (QSL)~\cite{Mandelstam1945, Anandan1990}. Even though, how fast a quantum system evolves in time was addressed in 
Ref. \cite{Mandelstam1945}, the notion of speed of transportation of state vector was formally defined using the Fubini-Study metric in Ref. \cite{Anandan1990}  and using the Riemannian metric in Ref. \cite{akp91}. Subsequently, an alternate speed limit for quantum state evolution was proved involving the average energy above the ground state of the Hamiltonian\cite{ Margolus1998}.  The QSL decides the minimal time of evolution of the quantum system. It entirely depends on intrinsic quantities of evolving quantum systems, such as the shortest path connecting the initial and final states of  the quantum system and the uncertainty in the Hamiltonian.

The QSL bounds were first investigated for the unitary dynamics of pure states
~\cite{Mandelstam1945,Anandan1990, Margolus1998,Levitin2009, Gislason1956, Eberly1973, Bauer1978, Bhattacharyya1983, Leubner1985,  Vaidman1992, Uhlmann1992, Uffink1993, Pfeifer1995,  Horesh1998, AKPati1999, Soderholm1999,  Andrecut2004, Gray2005, Luo2005,  Zielinski2006,  Andrews2007, Yurtsever2010, Fu2010, Zwierz2012, Poggi2013,Ness}. Later, QSL has been studied for the case of unitary dynamics of mixed states~\cite{Kupferman2008, Jones2010, Chau2010, S.Deffner2013, Fung2014, Andersson2014, D.Mondal2016, Mondal2016, S.Deffner2017, Campaioli2018}, unitary dynamics of multipartite systems~\cite{Giovannetti2004, Batle2005, Borras2006, Zander2007} and for more general dynamics \cite{Deffner2013, Campo2013, Taddei2013, Fung2013, Pires2016, S.Deffner2020,Jing2016,Pintos2021,Mohan,Hamazaki}. The study of QSL is significant for theoretical understanding of quantum dynamics and has relevance in developing quantum technologies and devices, etc. The QSL has applications in several fields, such as quantum computation~\cite{AGN12}, quantum thermodynamics~\cite{Mukhopadhyay2018,Funo2019}, quantum control theory~\cite{Caneva2009,Campbell2017}, quantum metrology~\cite{Campbell2018}, and so on.

In quantum physics, during the arbitrary dynamical evolution of the quantum system, we often encounter the situation where time evolved state becomes perfectly distinguishable (orthogonal) to initial state. At the same time, the expectation value of a given observable does not change or changes at a slower rate. For instance, let us consider a closed quantum system with internal Hamiltonian  $\sigma_{z}$ and initial state is $|+\rangle$, this initial state evolves to its orthogonal state $|-\rangle$ (up to a phase) by an external Hamiltonian  $\sigma_{y}$, where $|+\rangle$ and $|-\rangle$ are the eigenstates of $\sigma_{x}$. In this scenario, the initial state and final state of the system are distinguishable. However, both the initial and final states are energetically indistinguishable, so the evolution time for the average energy of the quantum system is zero. Similarly, one can consider that if a given quantum system interacting with a pure dephasing environment (dephasing in $\sigma_{z}$), then its state evolves to a decohered state. However, the expectation value of energy does not change in this process. The above discussion suggests that observables of a system can have different quantum speed limit bounds.

In the Schr\"{o}dinger picture, the state vector evolves in time, while in the Heisenberg picture, the observable of the quantum system evolves, and both of these formalisms are equivalent. In quantum mechanics, there is also interaction picture where both the state and the observables can change in time. This has important application in quantum field theory and many-body physics. In this paper we will use the Heisenberg picture for most of our discussions. The natural question, then, arises is how fast an observable evolves in time? Specifically, we will answer the question: how to obtain a lower bound on the evolution time of quantum observable and define the quantum speed limit for observable? Thus, seemingly technical difference between the Schr{\"o}dinger and the Heisenberg pictures, becomes rather important in the context of many-body physics. Here, often one cannot describe a state analytically due to its immense computational complexity. However, it is possible to compute expectation values of local observables in an efficient manner. Thus, it is desirable to bound how fast the expectation value corresponding to an observable changes in time in the Heisenberg picture.

Another motivation for studying observable speed limit is that this can have application in understanding the operator growth in many-body physics. In the context of complex systems, one of the pressing question is to understand the universal operator growth hypothesis~\cite{Perales2008}. 
The observables of a system which may be represented as operators in quantum
systems tend to grow over time, i.e., they become more complicated as the system evolves in time.
If we start with a simple many-body operator at some initial time, then because of the interaction Hamiltonian, the operator may 
become complex at a later time.
The quantum speed limit for the observable can answer a fundamental question on how fast a many body operator can tend to be complex?  
It is important know the rate of operator growth and we believe that the quantum speed limit for observable can throw new light on this question.

In complex quantum systems and many-body physics the bound on the commutator of
two operators, one operator being the time evolved version of an operator with support on some
region and the another operator with support on some other region,
plays a major role in the derivation of the Lieb-Robinson bound~\cite{Lieb,Hastings2010}. The later proves that 
the speed of propagation of perturbation in quantum system is bounded. 
Physically, this implies that for small times only small amounts of information can propagate 
from one region of the many-body system to another region.
While the Lieb-Robinson bound leads to a speed limit of information in quantum
systems, the quantum speed limit for the observable can answer the question how fast the commutator changes
for observables belonging to two distant regions. 
This is also important in the physical context where the underlying dynamics is highly chaotic \cite{gorin}.
The growth of the commutator between
two operators as a function of their separation in time has been used to quantify the rate of growth of chaos and 
information scrambling \cite{hay,sus,stan}. Therefore, the quantum speed limit for the commutator can answer the question how fast a
localized perturbation spreads in time in a quantum many-body system.
Since the time scale over which scrambling of information occurs is distinct from the
relaxation time of physical system, the observable quantum speed limit for the two-time commutator can play an important role 
in giving an estimate of the scrambling time in complex quantum systems.

To answer these fundamental questions, we formally introduce the notion of the quantum speed limit for observable. It is characterized as the maximal evolution speed of the expectation value of the given observable of the quantum system
during arbitrary dynamical evolution which can be unitary or non-unitary. It sets the bound on the minimum evolution time of quantum system required to evolve between different expectation values of a given observable. We do this for both closed and open quantum dynamics. We illustrate our main results for the ergotropy rate of quantum battery, rate of probability which also gives the standard QSL for state of the system. Moreover, we also compute the QSL for two-time correlation of an observable, which is a central quantity in the theory of quantum transport and complex quantum systems. We also apply our bound to obtain quantum speed limit for the commutator of two observables belonging to two distant regions in a many-body system. Our result can be equally important like the Lieb-Robinson bound.

\section{Observable Quantum Speed Limit (OQSL)}

Here, we show how to obtain QSL for observable.
This will answer the question how fast the expectation value of a given observable changes in time instead of the state of the quantum system. The observable QSL is defined as the maximum rate of evolution of the expectation value of an observable of a given quantum system during dynamical evolution.  It establishes a limit on the minimum evolution time necessary to evolve between different expected values of a given observable of the quantum system. Here, we will derive the observable QSL for closed and open system dynamics.

\subsection{Unitary Dynamics}
In this section, we will derive a bound for the observable undergoing unitary dynamics in the Heisenberg picture. Let us consider a closed quantum system whose initial state is $|\psi\rangle \in \cal{H}$ and its dynamical evolution dictated by unitary $U(t)=e^{-iHt/\hbar}$, where $H$ is time-independent driving Hamiltonian of the quantum system. Here, we want to determine how fast a quantum observable $O$ of the quantum system evolves in time and lower bound on its minimal evolution time. We know that the Heisenberg equation of motion governs the time evolution of an observable, which is given by
\begin{equation}\label{HB:eqn}
  i \hbar \frac{d O(t)}{dt}= [O(t), H],
\end{equation}
where  $H$ is the Hamiltonian of the system and $O(t) = U^{\dagger}(t)O(0)U(t)$ and $U^{\dagger}(t)U(t) = I$.

Now, we take the average of the above in the state $\ket{\psi}$ and take the absolute value of Eq \eqref{HB:eqn}. On using the Heisenberg-Robertson uncertainty relation, i.e., $\Delta A \Delta B \geq \frac{1}{2}\langle[A,B]\rangle$, where $A$ and $B$ are two incompatible observables~\cite{Robertson1929}, we obtain the following inequality
\begin{equation}\label{rate:1}
 \left| \frac{d\langle O(t)\rangle}{dt}\right|= \frac{1}{\hbar}| \langle[O(t), H]\rangle| \leq \frac{2 \Delta O(t) \Delta H  }{\hbar},
\end{equation}
where $\Delta O(t)$ = $\sqrt{\langle O(t)^{2} \rangle - \langle O(t)\rangle^2}$ and $\Delta H$ = $\sqrt{\langle H^{2} \rangle - \langle H \rangle^2 }$. 

The above inequality~\eqref{rate:1} is the upper bound on that the rate of change of expectation value of given observable of the quantum system evolving under unitary dynamics. 
 After integrating the above inequality with respect to time, we obtained the desired bound
\begin{equation}\label{Bound:1}
   T \geq \frac{\hbar}{2\Delta H}\int_{0}^{T}\frac{|d\langle O(t)\rangle|}{ \Delta O(t)},
\end{equation}
where,  we call the quantity (right hand side of inequality) $T^{O}_{QSL}=\frac{\hbar}{2\Delta H}\int_{0}^{T}\frac{|d\langle O(t)\rangle|}{ \Delta O(t)}$ quantum speed limit time of observable (OQSL), i.e., $T\geq T^{O}_{QSL}$.

If an observable $O$ satisfy the condition $O^{2}=I$, i.e., for the self-inverse observable, the above inequality can be expressed as
\begin{equation}\label{Bound:12}
   T \geq  \frac{\hbar}{2\Delta H}|\arcsin{\langle O(T) \rangle}-\arcsin{\langle O(0) \rangle}|.
\end{equation}

Here, we illustrate the tightness of the above speed limit for the observable with a simple example. Consider a qubit in a pure state $|\psi\rangle  = \alpha|0\rangle + \sqrt{1-\alpha^2}|1\rangle$ $(0 \leq  \alpha \leq 1)$ and it does not evolve in time.  The Hamiltonian of the system is given by $H$ = $\hat{m}.\Vec{\sigma}$, with $\hat{m}$ is unit vector. In the Heisenberg picture, let an observable evolves in time and we would like to evaluate the minimum evolution time of the expectation value of given observable $O(0)=\hat{n}.\Vec{\sigma}$, with $\hat{n}$ is unit vector in the state $|\psi\rangle $. We can calculate the following quantities $\Delta H=1$,
$\langle O(0) \rangle=1$
and $\langle O(T) \rangle=-1$, where we assume $\hat{n}= (1,0,0)$, $\hat{m}=(0,0,1)$, $\alpha=\frac{1}{\sqrt{2}}$ and $T = \frac{\pi}{2}$.
By using Eq \eqref{Bound:12}, we can obtain 
$T_{QSL}^{O}= \frac{\pi}{2} =T$. Hence, the bound given by Eq \eqref{Bound:12} is indeed tight. 
This simple example illustrates the usefulness of OQSL. The bound given in \eqref{Bound:1} may be thought of as the analog of the MT bound for the observable.

\textbf {\it Quantum speed Limit for States:}
Here, we discuss how the QSL for observable is connected to the standard QSL for state of a quantum system. We will show that the standard state-speed limit for the state, may be viewed as a special case of observable speed limits when the observable is chosen to be a projector on the initial state. 
Let us consider the initial state of a quantum system which is prepared in a state $|\psi\rangle=\sum_{i}a_{i}|i\rangle $. If we choose our observable to be a projector, i.e., $O(0)=P$, then the probability of finding quantum system in state $|i\rangle$ at $t=0$ is  $p(0)=|a_{i}|^2$ 
(if we measure a projector $P=|i\rangle \langle i|$). 
Here, we want to obtain the speed limit for the projector for the unitarily evolving quantum system, i.e., how fast the probability of finding quantum system in state
$|i\rangle$ changes in time. From \eqref{Bound:1}, then we can obtain the following inequality
\begin{equation*}
   T \geq \frac{\hbar}{2\Delta H}\int_{0}^{T}\frac{|d \langle P(t) \rangle|}{ \sqrt{\langle P(t) \rangle (1-\langle P(t) \rangle)} },
\end{equation*}
where $P(t)= U^{\dagger}(t)P(0)U(t)$ and $\langle P(t) \rangle$ = $p(t)$ is the probability of the quantum system in state $|i\rangle $ at a later time.
\begin{equation}
   T \geq  \frac{\hbar}{\Delta H}|\arcsin{ (\sqrt{p(T)} )}-\arcsin{ (\sqrt{p(0)} )}|.
\end{equation}

If $p(0) = 1$ i.e.  $|\psi\rangle=|i\rangle $ then above inequality yields the well known Mandelstam-Tamm bound of QSL for state evolution 
\begin{equation}\label{SQSL}
   T \geq  \frac{\hbar}{\Delta H}\arccos{ (\sqrt{p(T)} )}.
\end{equation}
This is the usual QSL obtained by Mandelstam-Tamm~\cite{Mandelstam1945} and Anandan-Aharaonov~\cite{Anandan1990}. Thus, the observable QSL also leads to standard QSL for the state change. In this sense, our approach also unifies the existing QSL.
Note that, this is an expression for survival probability and it is related to fidelity decay, which is an important quantity in quantum chaos~\cite{Herrera2014}.

Since the bound \eqref{Bound:1} is relatively harder to compute, one can derive the alternate bound for arbitrary initial state which may be pure or mixed (see Appendix \ref{app:B}), 
which is easier to compute.
\begin{equation}\label{Bound:2}
    T \geq \frac{\hbar}{2\sqrt{{\rm tr}(\rho^2)}}  \frac{|\langle O(T) \rangle  - \langle O(0)\rangle|}{\norm{O(0)H}_{\rm hs}},
\end{equation}
where ${\rm tr}(\rho^2)$ is the purity of the initial state, $\norm{A}_{\rm hs}$=$\sqrt{{\rm tr}{(A^{\dagger}A)}}$ is the Hilbert-Schmidt norm of operator $A$ and the RHS of the last equation is defined as $T^{O}_{QSL}$. The bound \eqref{Bound:2} suggests that if the initial state of 
the system is mixed, then observable evolves slower (OQSL depends on initial state). However, this bound \eqref{Bound:2} may not always be tight compared to the bound \eqref{Bound:1}. 

The above result can have an interesting application in physical systems where one try to estimate approximate conserved quantities. We know that if $O$ is the generator of a symmetry operation that acts on the physical system and if it commutes with the Hamiltonian, then it is conserved. Suppose that $O$ does not commute with the Hamiltonian. Then, we know that the observable will not be conserved. However, \eqref{Bound:2} can suggest that 
over some time interval $T$, how much $ \langle O(T) \rangle$  differs from $ \langle O(0) \rangle$. For a pure state, this difference is upper bounded by $\frac{2T}{\hbar} \norm{O(0)H}_{\rm hs} $.


Next, we obtain another QSL for observable. Let us consider a quantum system with a pure state $\rho=|\psi\rangle \langle \psi|$. The time evolution of expectation value any system observable $O$ is given as
\begin{equation}
 \langle O(t) \rangle={\rm tr}[U^{\dagger}(t)O(0)U(t)\rho].
\end{equation}
To find the rate of change of expectation value of observable, we need to differentiate the above equation with respect to time. By taking the absolute value of the rate equation, and apply the triangular inequality and the H\"older inequality~\cite{Rogers1888,Holder1889,Bhatia1997}, we obtain the following inequality 

\begin{equation}\label{rate:2}
\left|\frac{ d\langle O(t) \rangle}{dt}\right|
  \leq\frac{2}{\hbar}\norm{HO(t)}_{\rm op}.
 \end{equation}
The above inequality~\eqref{rate:2} provides the upper bound on that the rate of change of expectation value of given observable of the quantum system evolving under unitary dynamics.

The operator norm,  the Hilbert-Schmidt norm and the trace norm of an operator satisfy the inequality $\norm{A}_{\rm op} \leq \norm{A}_{\rm hs}   \leq \norm{A}_{\rm tr} $ and 
the operator norm is unitary invariant  $\norm {U^{\dagger}AU}_{\rm op}$ = $\norm {A}_{\rm op}$, then we can obtain the following bound as (see Appendix \ref{app:C})
\begin{equation}\label{Bound:3}
    T \geq \frac{\hbar}{2}   \{\frac{|\langle O(T) \rangle  - \langle O(0)\rangle| }{min\{\norm{O(0)H}_{\rm op},\norm{O(0)H}_{\rm tr}\}}\},
\end{equation}
where $T^{O}_{QSL}$= $\frac{\hbar}{2}  \{\frac{|\langle O(T) \rangle  - \langle O(0)\rangle| }{min\{\norm{O(0)H}_{\rm op},\norm{O(0)H}_{\rm tr}\}}\}$  quantum speed limit time of observable (OQSL).

One should consider the maximum of \eqref{Bound:1}, \eqref{Bound:2} and \eqref{Bound:3} for the tighter bound. For the unitary evolution, the bounds \eqref{Bound:1}, \eqref{Bound:2} and \eqref{Bound:3} determines how fast the expectation value of an observable of the quantum system changes in time. If a given observable's initial and final expectation value does not change undergoing unitary evolution, then OQSL is zero. Since the minimal evolution time for state evolution cannot be zero in the above scenario, the aforementioned scenario is the major difference between OQSL bounds and standard QSL bounds (MT and ML bounds) for unitary dynamics.

\subsection{Arbitrary Dynamics}

In general, for an arbitrary observable $O$ one can obtain the following inequality using the triangle inequality for the absolute value, and the H\"{o}lder inequality (see Appendix \ref{app:D})
\begin{equation}
|\langle O \rangle_{\rho} - \langle O \rangle_{\sigma}|\leq 2\norm{O}_{\rm op}l(\rho,\sigma),
\end{equation}
where $l(\rho,\sigma)$ = $\frac{1}{2} {\rm tr }\lvert \rho -\sigma \rvert$ is the trace distance between states $\rho$ and $\sigma$.

With the help of the above inequality relation, we can define a distance that captures the change in the expectation of an observable during the arbitrary dynamical evolution (in general the evolution governed by the master equation $\dot{O}(t)=L_{t}^{\dagger}(O(t))$, where ${L}^{\dagger}_{t}$ is adjoint of the Liouvillian super-operator, which can be unitary or non-unitary), and it is given by

\begin{equation}\label{OD}
\cal{D}(O(t),O(0))   = \frac{|{\rm tr}(O(t)  -  O(0) ){\rho})|}{2\norm{O(0)}_{\rm op}},
\end{equation}
 where $\norm{O(0)}_{\rm op}$ is a re-scaling factor because the spectral gap in the observable can be arbitrarily large.

Using Eq \eqref{OD} we can obtain the desired QSL bound on evolution time of expectation value of an observable for arbitrary dynamics as
\begin{equation}\label{arbitrary}
T \geq  \frac{|\langle O(T) \rangle  -  \langle O(0) \rangle|}{\sqrt{{\rm tr}(\rho ^2)}\Lambda_{T}}
\end{equation}
where $\Lambda_{T}=\frac{1}{T}\int_{0}^{T}dt{\norm{{L_{t}^{\dagger}({O}(t))}}_{\rm hs}}$ is evolution speed of observable $O$ and $T^{O}_{QSL}=\frac{|\langle O(T) \rangle  -  \langle O(0) \rangle|}{\sqrt{{\rm tr}(\rho ^2)}\Lambda_{T}}$ quantum speed limit time of observable (OQSL).

Details of the derivation are provided in Appendix \ref{app:E}.
For arbitrary dynamics, the bound \eqref{arbitrary} determines how fast the expectation value of an observable of the quantum system changes in time. The derived bound \eqref{arbitrary} implies that the OQSL is dependent on the purity of the initial state of the quantum system as well as the evolution speed of the observable.

\subsection{Lindblad Dynamics}
Let us consider a quantum system (S) that is interacting with its environment (E). The  total Hilbert space of combined system is $\cal{H}_{S}$ $\otimes$ $\cal{H}_{E}$ and  we  assume  the initial  state of combined system is  represented  by  the  separable  density  matrix  $\rho_{SE}(0)$ $=$ $\rho$ $\otimes$ $\sigma$, where quantum system's initial state $\rho$ can be pure or mixed and $\sigma$ is state of environment. The Lindbladian ($\mathcal{L}$) governs the reduced dynamics of a quantum system (S). Here, we aim to determine how fast the expectation value of given observable $O$ of the reduced quantum system (S) evolves in time and lower bound on its minimal evolution time. Let the quantum system has an internal Hamiltonian $H_{S}$. If the system interacts with its surroundings, then, the dynamics of given observable of the quantum system is governed by the Lindblad master equation in the Heisenberg picture. Hence, the expectation value of the observable $O$ belonging to the system follows the dynamics
\begin{equation}
 \langle O(t)\rangle  ={\rm tr}(O(0)\Phi_{t}(\rho)) = {\rm tr}(\Phi^{\dagger}_{t}(O(0))\rho),
\end{equation}
where $\Phi_{t}$ is generator of dynamics, $O(t)= \Phi_{t}(O(0))=e^{\mathcal{L}^{\dagger}t}O(0)$
and ${\mathcal{L}^{\dagger}}$ is adjoint of Lindbladian. 

The time evolution of a observable $O$ is given by the following equation
\begin{equation}\label{Lind}
\frac{d O(t)}{dt} =   {\mathcal{L}^{\dagger}}[O(t)],
\end{equation}
where $\mathcal{L}^{\dagger}[O(t)]=\frac{i}{\hbar}[H_{S},O(t)] + D[O(t)]$, $D[O(t)]=\sum_{k}\gamma_{k}(t)(L_{k}^{\dagger}O(t)L_{k}-\frac{1}{2}\{L_{k}^{\dagger}L_{k},O(t)\})$ and $L_{k}$'s are jump operators of the system.

Let us take the average of Eq~\eqref{Lind} in the state $\rho$ and its absolute value. By applying the Cauchy-Schwarz inequality, we obtain the following inequality

\begin{equation}\label{rate:4}
\left|\frac{d\langle O(t) \rangle}{dt}\right| \leq \sqrt{{\rm tr}(\rho^{2})}\norm{{\mathcal{L}^{\dagger}}[O(t)]}_{\rm hs},
\end{equation}
where $\norm{A}_{\rm hs}$ = $\sqrt{{\rm tr}(A^{\dagger}A)}$ is the Hilbert-Schmidt norm of operator A.

The above inequality~\eqref{rate:4} is the upper bound on that the rate of change of expectation value of given observable of the quantum system evolving under Lindblad dynamics.
After integrating the above inequality, we obtained the following bound

\begin{equation}\label{Bound:open1}
    T \geq   \frac{|\langle O(T) \rangle-\langle O(0) \rangle|}{\sqrt{{\rm tr}(\rho^2)}\Lambda_{T}},
\end{equation}
where $\Lambda_{T}=\frac{1}{T}\int_{0}^{T}dt\norm{{\mathcal{L}^{\dagger}}[O(t)]}_{\rm hs}$ is evolution speed of the observable $O$ and $T^{O}_{QSL}$= $\frac{|\langle O(T) \rangle-\langle O(0) \rangle|}{\sqrt{{\rm tr}(\rho^2)}\Lambda_{T}}$  quantum speed limit time of observable (OQSL).

For the Lindblad dynamics, the bound given in \eqref{Bound:open1} determines how fast the expectation value of an observable of the quantum system changes in time. The obtained bound  \eqref{Bound:open1} suggests that the OQSL depends on the purity of the initial state of the evolving quantum system. Note that, OQSL time is zero if the expectation value of a given observable does not change during the dynamics.

\textit{\bf Comparison between QSL and OQSL for pure dephasing dynamics:}
Let us consider the QSL bound for open quantum systems governed by a Lindblad quantum master equation. For a Markovian dynamics of an open quantum
system expressed via a Lindbladian ($\mathcal{L}$), the following lower bound on the evolution time needed for a quantum system to evolve from initial state $\rho_0$ to final state $\rho_T$ was given in Ref.~\cite{Campo2013} as

\begin{equation}\label{qslbound:1}
    T \geq \frac{|\cos{\theta}-1|{\rm tr}[\rho_0^2]}{\sqrt{{\rm tr}[\mathcal{L}^\dagger(\rho_0)]^2}}
\end{equation}

where $\theta=\cos^{-1}(\frac{{\rm tr}[\rho_0 \rho_T]}{{\rm tr}[\rho_0^2]})$ is expressed in terms of relative purity between the initial and the final state.

Let us consider a two-level quantum system with the ground state $|1\rangle\langle 1|$  and the excited state $|0\rangle\langle 0|$ interacting with a dephasing bath. The corresponding dephasing Lindblad or jump operator of the system is given by $L_{0} = \sqrt{\frac{\gamma}{2}} \sigma_{z}$, where  $\sigma_{z}$ is the Pauli-$Z$ operator and $\gamma$ is a real parameter denoting the strength of dephasing. The Lindblad master equation~\cite{Lindblad1976} governs the time evolution of two-level quantum system, and it is given by
\begin{equation}
\frac{{ d} \rho_t}{{ d} t} = \cal{L}(\rho_t)= \frac{\gamma}{2}(\sigma_{z}\rho_t\sigma_{z}-\rho_t).
\end{equation}
If the quantum system initially in a state  $\rho_{0}=|+\rangle \langle +|$, where $|+\rangle = {\frac{1}{\sqrt{2}}}(|0\rangle + |1\rangle)$, then solution of the Lindblad equation is given by
\begin{align}
\rho_t=\frac{1}{2} (|0\rangle \langle 0| + |1\rangle \langle 1| 
+ e^{-\gamma t} (|1\rangle \langle 0|+ |0\rangle \langle 1|)).
\end{align}
If given  observable is $O(0)$ = $\sigma_{x}$, then the solution of~\eqref{Lind} for the dephasing dynamics is given as
\begin{equation}
    O(t)= e^{- \gamma t} \sigma_{x}
\end{equation}

To estimate bounds \eqref{Bound:open1} and \eqref{qslbound:1}, we require the following quantities:
\begin{equation}
    {\rm tr}[\rho_0^2]=1
\end{equation}

\begin{equation}
   \cos{\theta} =\frac{1+e^{-\gamma  t}}{2}
\end{equation}
\begin{equation}
    {\rm tr}[\mathcal{L}^\dagger(\rho_0)]^2=\frac{\gamma ^2}{2}
\end{equation}

\begin{equation}
    \langle O(0) \rangle = {\rm tr}[\sigma_{z}\rho_0]=1
\end{equation}

\begin{equation}
    \langle O(t) \rangle = {\rm tr}[e^{-\gamma t}\sigma_{z}\rho_0]=e^{-\gamma t}
\end{equation}

\begin{equation}
    \norm{{\mathcal{L}^{\dagger}}[O(t)]}_{\rm hs}=\sqrt{2}\gamma e^{- \gamma t} 
\end{equation}

\begin{figure}
    \centering
    \includegraphics[width=9cm]{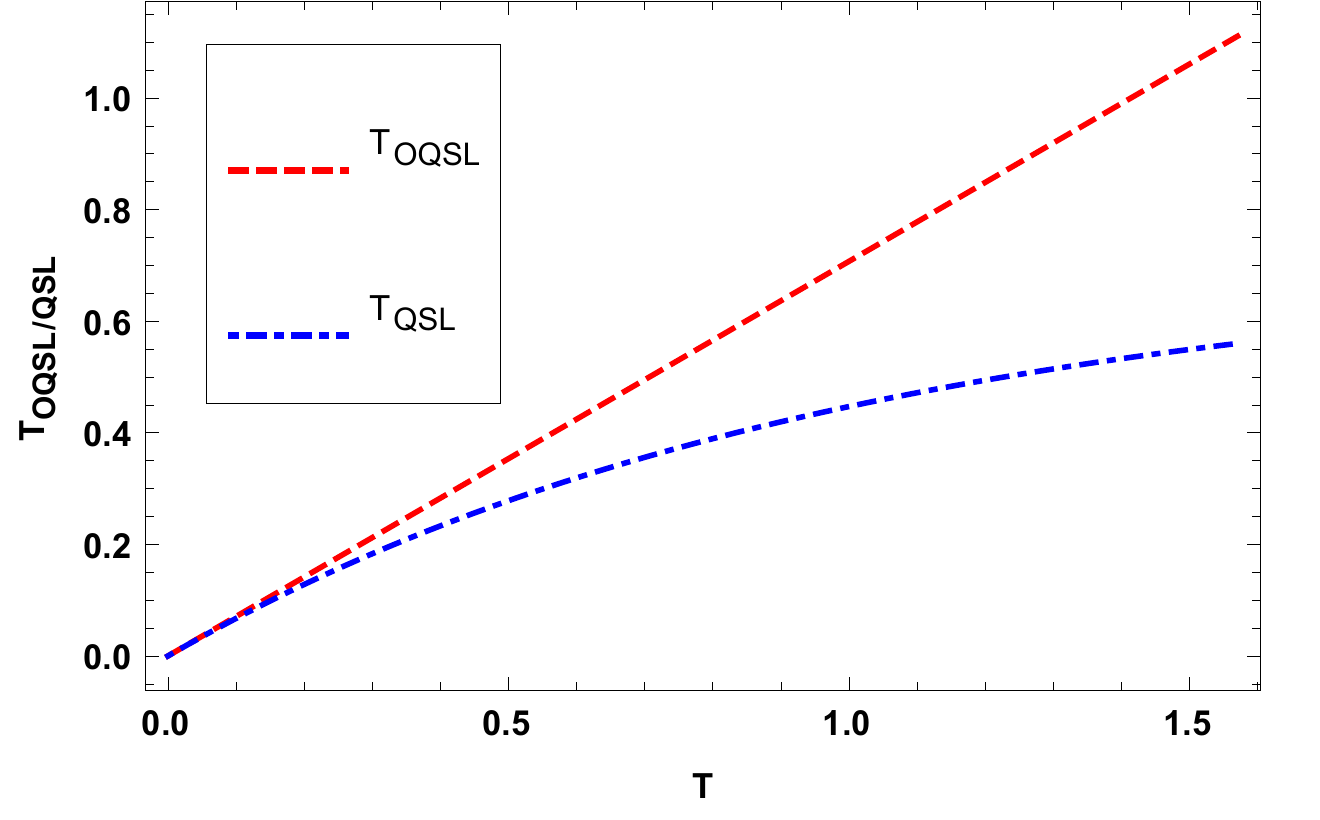}
    \caption{Here we depict $T_{QSL/OQSL}$ vs $T$ and we have considered $\gamma=1$.}
    \label{fig}
\end{figure}

In Fig.~\ref{fig}, we plot $T_{QSL/OQSL}$ vs $T$ $\in$ $[0, \frac{\pi}{2}]$ for pure dephasing dynamics and we have considered $\gamma=1$. Fig.~\ref{fig} shows that our OQSL bound \eqref{Bound:open1} is tighter than QSL bound \eqref{qslbound:1} for pure dephasing process. Both bounds (our bound \eqref{Bound:open1} and bound \eqref{qslbound:1}) are obtained by employing the Cauchy-Schwarz inequality. Therefore, one expects that both these bounds to be equally tight, but it is not true. It turns out that bound \eqref{qslbound:1} is loose. It happens because while deriving the bound \eqref{qslbound:1} in Ref.~\cite{Campo2013}  the authors use an additional inequality along with the Cauchy-Schwarz inequality, i.e., ${\rm tr}(\rho_t^2) \leq 1$ (see Eq (7) of Ref.~\cite{Campo2013}). They did this to obtain a time-independent bound on the rate of change of the purity.

\subsection{Dynamical Map}
We can also express the QSL for the observable using the Kraus operator evolution. If a given quantum system has initial state $\rho$, and let its dynamical evolution is governed by a CPTP map ($\mathcal{E}$)
which is described by a set of Kraus operators $\{K_{i}(t)\}$ and $\sum_{i}K^{\dagger}_{i}(t)K_{i}(t)=\mathcal{I}_{S}$. 
 The dynamics of the observable $O$ in the Heisenberg picture is described as
\begin{equation}\label{OE}
 O(t) =\sum_{i}K^{\dagger}_{i}(t)O(0)K_{i}(t).
\end{equation}

Using Eq \eqref{OE}, we obtain the QSL for the observable as given by

\begin{equation}\label{Bound:open2}
    T \geq   \frac{|\langle O(T) \rangle-\langle O(0) \rangle|}{2\sqrt{{\rm tr}(\rho^2)}\Lambda_{T}},
\end{equation}
where $\Lambda_{T}=\frac{1}{T}\int_{0}^{T}dt\norm{{K}^{\dagger}_{i}(t)O(0)\dot{K}_{i}(t)}_{\rm hs}$ is the evolution speed of the observable $O$ and $T^{O}_{QSL}$= $\frac{|\langle O(T) \rangle-\langle O(0) \rangle|}{2\sqrt{{\rm tr}(\rho^2)}\Lambda_{T}}$ quantum speed limit time of observable (OQSL).

Details of the derivation are provided in Appendix \ref{app:F}.
For dynamical map dynamics, the bound \eqref{Bound:open2} determines how fast the expectation value of an observable of the quantum system changes in time. According to the obtained bound \eqref{Bound:open2} the OQSL depends on the purity of the initial state of evolving quantum system and on the speed of observable evolution.

\subsection{State independent QSL for observables }

The bounds for QSL for observable that have been proved in previous sections are state-dependent bounds. One may be
curious to know if we can prove some state-independent bounds, i.e.,
whether we can derive the bounds which are given merely in terms of properties
of the observables themselves.  Here, we make an
attempt to formulate a bound without optimising over states.
To derive the state independent speed limit for observable, consider the Hilbert-Schmidt inner product for observables. 
The Hilbert-Schmidt inner product of two observables $O(0)$ and $O(t)$ is defined as

\begin{equation}
    \langle O(0), O(t) \rangle = {\rm tr}[O(0)O(t)],
\end{equation}
where $O(t)=e^{{L}^{\dagger}t}O(0)$ (${{L}^{\dagger}}$ is adjoint of Liouvillian super-operator).

After differentiating above equation with respect to time, we obtain
\begin{equation}
   \frac{d}{dt}\langle O(0), O(t) \rangle = {\rm tr}[ O(0)\dot{O}(t) ] ={\rm tr}[O(0)L^{\dagger}({O}(t))].
\end{equation}
Let us take the absolute value of the above equation. Then by applying the Cauchy-Schwarz inequality, we can obtain the following inequality 

\begin{equation}
      \left|\frac{d}{dt}\langle O(0), O(t) \rangle\right| \leq \norm{O(0)}_{\rm hs}\norm{L^{\dagger}(O(t))}_{\rm hs}.
\end{equation}
After integrating the above inequality, we obtain the following bound
\begin{equation}\label{StateINQSL}
T \geq   \frac{ |\langle O(0), O(T) \rangle - \langle O(0), O(0) \rangle|}{ \norm{O(0)}_{\rm hs}\Lambda_{T}},
\end{equation}
where $\Lambda_{T}=\frac{1}{T}\int_{0}^{T}dt\norm{{{L}^{\dagger}}[O(t)]}_{\rm hs}$ is the evolution speed of the observable $O$ and $T^{O}_{QSL}= \frac{ |\langle O(0)|O(T) \rangle - \langle O(0)|O(0) \rangle|}{ \norm{O(0)}_{\rm hs}\Lambda_{T}}$ quantum speed limit time of observable (OQSL).

The bound \eqref{StateINQSL} is independent of state of the quantum system and it is applicable for arbitrary dynamics which can be unitary or non-unitary.
In future, it will be worth exploring if it is possible to obtain some bounds using different approach, for example, which may involve
separation between extreme eigenvalues of the operators or optimising over possible initial states. The state-independent bound will have its own merit
as they could be understood as representing some best case scenario where the
time is shortest possible to modify the expectation value of a given
observable when optimizing over all possible states. These kind of bounds may find applications in the context of quantum metrology, where we
optimize over the probe states in order to obtain the “fastest change”
of the state from the point of view of certain parameters.

\section{Applications}
In this section, we illustrate the usefulness of observable QSL for quantum battery, growth of two-time correlation function and connection to the Lieb-Robinson bound.  
\subsection{Quantum batteries}
A {\it Quantum battery} is a microscopic energy storage device introduced by R. Alicki, and M. Fannes \cite{Alicki2013}. Several theoretical works have been done to strengthen this novel idea of quantum battery and enhance its non-classical features \cite{Binder2015,Campaioli2017,Ferraro2018,F.Campaioli2018,Julia2020,Andolina2019,G.Andolina2019,Andolina2018,Le2018,S.Ghosh2020,Luis2020,Crescente2020,S.Gherardini2020,A.C.Santos2019,Mohan2021}. Quantum batteries can easily outperform classical batteries because of several quantum advantages. Here, our main aim is to obtain a minimal unitary charging time of the quantum battery using the observable QSL.

 The quantum battery consists of many quantum systems with several degrees of freedom in which we can deposit or extractwork from it. Let us consider the battery with Hamiltonian $H_{B}$, and it is charged by field $H_{C}$. 
The total Hamiltonian of the quantum battery is described by
\begin{equation}
    H_{T}= H_{B} + H_{C}.
\end{equation}
The amount of extractable energy from the quantum system by unitary operations
is termed the ergotropy of quantum battery \cite{F.Campaioli2018}. Which is given by
\begin{equation}\label{ergotropy}
    \varepsilon(t) = \langle \psi(t)|H_{B}| \psi(t) \rangle -  \langle \psi(0)|H_{B}| \psi(0) \rangle,
\end{equation}
where $|\psi(0)\rangle$ and $|\psi(t)\rangle$ are initial and final state of quantum battery while charging.

Note that the above expression holds true in the Schr\"odinger picture. However, in the Heisenberg picture
the ergotropy can be rewritten as 
\begin{equation*}
    \varepsilon(t) = \langle \psi(0)|(H_{B}(t)-H_{B}(0))| \psi(0) \rangle,
\end{equation*}
where $H_{B}(t)$ = $ e^{iHt/\hbar} H_{B}(0)e ^{-iHt/\hbar}$ and $H_{B}(0)$ = $H_{B}$.

The rate of change ergotropy of quantum battery during the charging process can be obtained by differentiating the above equation with respect to time, which is given by
\begin{equation*}
   \frac{ d \varepsilon(t)}{dt}  =\frac{d}{dt}\langle \psi(0)|H_{B}(t)| \psi(0) \rangle.
\end{equation*}
Using our bound we can write the QSL for ergotropy as

\begin{equation}\label{CT1}
   T \geq \frac{\hbar}{2\Delta H_{T}}\int_{0}^{T}\frac{|d\langle H_{B}(t)-H_{B}(0) \rangle|}{ \Delta H_{B}(t)},
\end{equation}
where $T$ is charging time period of the quantum battery.

Also, an alternative unified bound we can be obtained by using bounds \eqref{Bound:2} and \eqref{Bound:3},

\begin{equation}\label{CT2}
    T \geq \frac{\hbar}{2}  \frac{|\langle H_{B}(T) \rangle  - \langle H_{B}(0)\rangle| }{\min\{\norm{\bullet}_{\rm op},\norm{\bullet}_{\rm hs},\norm{\bullet}_{\rm tr}\}},
\end{equation}
 where  $\bullet$ stands for ${H_{B}(0)H_{T}}$ and the operator norm,  the Hilbert-Schmidt norm and the trace norm of an operator satisfy the inequality $\norm{A}_{\rm op} \leq \norm{A}_{\rm hs}   \leq \norm{A}_{\rm tr} $.
 
Since previously obtained bounds~\cite{Campaioli2017,F.Campaioli2018} on charging time of quantum battery are based on distinguishability of the initial and final state vectors of the quantum battery, the bounds we have presented in this section are based on the difference of initial ergotropy and final ergotropy of the quantum battery. The bounds obtained in this section can easily outperform previously obtained bounds, especially when the battery has degenerate energy levels.

For example, let us consider the model of qubit quantum battery which has Hamiltonian $H_{B}=\sigma_{z}$. Let us consider battery is initially in state $|\phi^{+}\rangle= a|0\rangle +b |1\rangle$ (which has nonzero ergotropy) then by applying some charging field $H_{C}(t)$ we reached the final state $|\phi^{-}\rangle= a|0\rangle -b |1\rangle$. In this process neither we extracted any work from quantum battery nor stored any work in the quantum battery because both initial and final states have same ergotropy according to Eq \eqref{ergotropy}. Note that, if we calculate charging time according to standard QSL \eqref{SQSL} or bounds presented in ~\cite{Campaioli2017,F.Campaioli2018} we obtain nonzero minimal charging time but according to our bounds \eqref{CT1} and \eqref{CT2} minimal charging time is zero. This happens because standard QSL and bounds presented in ~\cite{Campaioli2017,F.Campaioli2018} are based on the notion of state distinguishability while our bounds \eqref{CT1} and \eqref{CT2} depends on change in the ergotropy. Therefore, our bounds \eqref{CT1} and \eqref{CT2} yields correct minimal charging time.

\subsection{Transport properties}
A crucial quantity in the theory of quantum transport in many body physics is the two-time correlation function of an observable. This section aims to obtain a speed limit for a two-time correlation of an observable and its time evolved observable. For an arbitrary pure quantum state $\rho$, we can define the two-time correlation function $C(A(t),A(0))$ between observables $A(t)$ and $A(0)$ as
\begin{equation}
    C(t) = \langle A(t)A(0) \rangle - \langle A(t) \rangle \langle A(0) \rangle.
\end{equation}

For the closed dynamics case $A(t)$ = $U^{\dagger}(t)A(0)U(t)$ ($U(t)=e^{-iHt/\hbar}$) and for the open dynamics case  $A(t)=e^{\mathcal{L}^{\dagger}t}A(0)$ (${\mathcal{L}^{\dagger}}$ is adjoint of Lindbladian).
We can derive the following speed limit bound on two-time correlation function for closed dynamics
\begin{equation}\label{QSLTTU}
 T \geq \frac{\hbar}{2} \frac{ |C(T) - C(0)|}{ \norm{ A(0)}_{\rm op} \frac{1}{T}\int_{0}^{T}dt\norm{[H,A(t)] }_{\rm op}}.
\end{equation}

Similarly, we can derive the following speed limit bound on two-time correlation function for open dynamics
\begin{equation}\label{QSLTTO}
 T \geq \frac{\hbar}{2} \frac{ |C(T) - C(0)|}{ \norm{ A(0)}_{\rm op} \frac{1}{T}\int_{0}^{T}dt\norm{{\mathcal{L}^{\dagger}}[A(t)]}_{\rm op}}.
\end{equation}

Details of the derivation of bounds \eqref{QSLTTU} and \eqref{QSLTTO} are provided in Appendix \ref{app:G} and \ref{app:H}.

\subsection{Relation to Lieb-Robinson bound}
The Lieb-Robinson bound~\cite{Lieb,Hastings2010} provides the speed limit for information propagation about the perturbation. This gives an upper bound for the operator norm of the commutator of $A(t)$ and $B$, where $A$ and $B$ are spatially separated operators of a many body quantum system. 
This bound implies that even in the 
non-relativistic quantum dynamics one has some kind of locality
structure analogous to the notion of finiteness of the speed of light in the relativistic theory.

This section aims to derive distinct speed limit bound for the commutator of $A(t)$ and $B$, i.e., how fast the commutator changes in the 
Heisenberg picture. The commutator of two observables in two different regions of a many-body ssytem is defined as
\begin{equation}
   O(t) = [B(0), A(t)].
\end{equation}
The average of the commutator in the state $\rho$ is given by $ \langle{O(t)}\rangle= {\rm tr}(O(t)\rho)$ and $\rho$ is pure state of given quantum system.

Here, we want to obtain the speed limit bound for commutator for closed system dynamics and open system dynamics both. For a closed dynamics $A(t)$ = $U^{\dagger}(t)A(0)U(t)$ ($U(t)=e^{-iHt/\hbar}$) and for open dynamics $A(t)=e^{\mathcal{L}^{\dagger}t}A(0)$ (${\mathcal{L}^{\dagger}}$ is adjoint of Lindbladian).
We can derive the following speed limit bound on the commutator for closed dynamics
\begin{equation}\label{CQSLU}
T \geq  \frac{2}{\hbar} \frac{ |\langle{O(T)}\rangle|}{ \norm{B(0)}_{\rm op}\frac{1}{T}\int_{0}^{T}dt\norm{[H,A(t)]}_{\rm op}}.
\end{equation}

Similarly, we can derive the following speed limit bound on the commutator for open dynamics 
\begin{equation}\label{CQSLO}
T \geq  \frac{2}{\hbar} \frac{ |\langle{O(T)}\rangle|}{ \norm{B(0)}_{\rm op}\frac{1}{T}\int_{0}^{T}dt\norm{{\mathcal{L}^{\dagger}}[A(t)]}_{\rm op}}.
\end{equation}
Details of the derivation of bounds \eqref{CQSLU} and \eqref{CQSLO} are provided in Appendix \ref{app:I} and \ref{app:J} .
Note that our bounds are state-dependent while the Lieb-Robinson bound is state-independent.
Also, to prove the Lieb-Robinson bound one needs bounded interactions such as those encountered in 
quantum spin systems, whereas the quantum speed limit for the commutator does not require any assumption about the underlying Hamiltonian.

\section{Conclusions}
The standard quantum speed limit for evolution of a state plays an important role in quantum theory, quantum information, quantum control and quantum thermodynamics. However, if we describe the quantum dynamics in the Heisenberg picture, then we cannot use the QSL for the state evolution. We need to define the evolution speed of observable for a quantum system in the Heisenberg
picture which has never been addressed before.
In this paper, we have derived the quantum speed limits for general observables for the unitary, the Lindbladian dynamics 
as well as completely positive dynamics for the first time. Along with this, we have presented several possible applications of these bounds such as in the quantum battery, probability dynamics, growth of two-point correlation function, time development of commutator and its connection to the Lieb-Robinson bound. A salient outcome of our approach is that the standard QSL for state can be viewed as a special case of QSL for observable. In future, 
we hope that these bounds can have useful applications 
in quantum metrology, quantum control, detection of non-Markovianity, quantum thermodynamics, charging and discharging of quantum batteries and many other areas as well.

\section{ACKNOWLEDGMENTS}
The research of BM was partly supported by the INFOSYS scholarship. We thank Kavan Modi for useful discussions and comments.

\bibliography{oqsl}

\appendix

\section{Derivation of Eq \eqref{Bound:2}}\label{app:B}
To obtain an alternate OQSL given in Eq \eqref{Bound:2}, let the state of a quantum system is described by a density operator $\rho$ (not necessarily pure). The time evolution of the expectation value any system observable $O$ is given as
\begin{equation}
 \langle O(t) \rangle= {\rm tr}[U^{\dagger}(t)O(0)U(t)\rho].
\end{equation}
After differentiating above equation with respect to time, we obtain 

\begin{equation*}
\begin{split}
  \frac{ d\langle O(t) \rangle}{dt}= {\rm tr}[\dot{U}^{\dagger}(t)O(0)U(t)\rho] + {\rm tr}[U^{\dagger}(t)O(0)\dot{U}(t)\rho].
   \end{split}
\end{equation*}

Let us take absolute value of the above equation and using the triangular inequality $|A+B|\leq|A|+|B|$, we obtain
\begin{equation*}
\begin{split}
\left|\frac{ d\langle O(t) \rangle}{dt}\right|
  &\leq\lvert {\rm tr}[\dot{U}^{\dagger}(t)O(0)U(t)\rho]\rvert \\
  &+ \lvert {\rm tr}[U^{\dagger}(t)O(0)\dot{U}(t)\rho]\rvert.
  \end{split}
\end{equation*}

Now, using the Cauchy-Schwarz inequality $|{\rm tr}(AB)|$ $\leq$ $\sqrt{{\rm tr}(A^{\dagger}A){\rm tr}(B^{\dagger}B)}$, we can obtain the following inequality

\begin{equation*}
|\frac{ d\langle O(t) \rangle}{dt} |
  \leq   2\sqrt{{\rm tr}[\dot{U}^{\dagger}(t)O^{2}(0)\dot{U}(t)]{\rm tr}({\rho}^2)}.
\end{equation*}

The above inequality can be further simplified as 
\begin{equation*}
   \left| \frac{ d\langle O(t) \rangle}{dt}\right|\leq \frac{2}{\hbar}\sqrt{{\rm tr}(\rho)^2} \norm {O(0)H}_{\rm hs},
\end{equation*}
where $\norm{A}_{\rm hs}=\sqrt{{\rm tr}(A^{\dagger}A)}$ is the Hilbert-Schmidt norm of operator A.

After integrating with respect to time, we obtained the following bound
\begin{equation}
    T \geq \frac{\hbar}{2\sqrt{{\rm tr}(\rho^2)}}  \frac{|\langle O(T) \rangle  - \langle O(0)\rangle|}{\norm{O(0)H}_{\rm hs}}.
\end{equation}

If an observable satisfy $O^{2}$ = $I$, then for pure state case the above bound can be expressed as
\begin{equation}
    T \geq \frac{\hbar}{2 \norm{H}_{\rm hs} }  |\langle O(T) \rangle  - \langle O(0)\rangle|
\end{equation}

This completes the proof of Eq \eqref{Bound:2}.
\section{Derivation of Eq \eqref{Bound:3}}\label{app:C}

To derive the bound given in Eq \eqref{Bound:3}, let us assume that a quantum system has state $\rho$ (pure state). The time evolution of the expectation value any system observable $O$ is given as
\begin{equation}
 \langle O(t) \rangle={\rm tr}[U^{\dagger}(t)O(0)U(t)\rho].
\end{equation}
To find the rate of change of expectation value of observable, we need to differentiate the above equation with respect to time, which is given by
\begin{equation*}
\begin{split}
\frac{ d\langle O(t) \rangle}{dt}
  = {\rm tr}[\dot{U}^{\dagger}(t)O(0)U(t)\rho] +  {\rm tr}[U^{\dagger}(t)O(0)\dot{U}(t)\rho].
  \end{split}
\end{equation*}

Let us take the absolute value of the above equation. Then by applying the triangular inequality $|A+B|\leq|A|+|B|$, we can obtain the following inequality 
\begin{equation}
\begin{split}
\left|\frac{ d\langle O(t) \rangle}{dt}\right|
  \leq\frac{1}{\hbar}(\lvert {\rm tr}[HO(t)\rho]\rvert + \lvert {\rm tr}[O(t)H\rho]\rvert).
  \end{split}
\end{equation}

Next, we use the H\"older inequality $|{\rm tr}(AB)|$ $\leq$ $\norm{A}_{\rm p}\norm{B}_{\rm q}$, where p, q $\in$ $[1,\infty)$ such that $\frac{1}{p}+\frac{1}{q}=1$  \cite{Rogers1888,Holder1889,Bhatia1997}. This leads to the following inequality 
\begin{equation*}
\begin{split}
\left|\frac{ d\langle O(t) \rangle}{dt}\right|
  \leq\frac{2}{\hbar}\norm{HO(t)}_{\rm op}.
  \end{split}
\end{equation*}

We know that the operator norm,  the Hilbert-Schmidt norm and the trace norm of an operator satisfy the inequality $\norm{A}_{\rm op} \leq \norm{A}_{\rm hs}   \leq \norm{A}_{\rm tr} $ and 
the operator norm is unitary invariant  $\norm {U^{\dagger}AU}_{\rm op}$ = $\norm {A}_{\rm op}$, then we can express the above inequality as
\begin{equation*}
\begin{split}
\left|\frac{ d\langle O(t) \rangle}{dt}\right|
  \leq  \frac{2}{\hbar}\norm{HO(0)}_{\rm tr}.
  \end{split}
\end{equation*}

After integrating the above inequality, we obtained the desired bound
\begin{equation*}
    T \geq  \frac{\hbar}{2}  \frac{|{ \langle O(T) \rangle -\langle O(0)\rangle}|}{\hspace{1mm}\norm{HO(0)}_{\rm tr}}.
\end{equation*}
In general, we can write the above bound as
\begin{equation}
    T \geq \frac{\hbar}{2}   \{\frac{|\langle O(T) \rangle  - \langle O(0)\rangle| }{min\{\norm{O(0)H}_{\rm op},\norm{O(0)H}_{\rm tr}\}}\}.
\end{equation}

This completes the proof of Eq \eqref{Bound:3}.

\section{Trace distance Bounds on Observable difference}\label{app:D}
We know that, we can use the trace distance to figure out how close two density operators are. Similar question we can ask for the expectation value of the observable.

Note that
 \begin{equation*}
\begin{split}
     |\langle O \rangle_{\rho} - \langle O \rangle_{\sigma}|& \equiv |{\rm tr}(\rho - \sigma)O| \leq {\rm tr}|(\rho - \sigma)O|= \norm{(\rho - \sigma)O}_{\rm tr} \\
  & \stackrel{\text{H\"{o}lder}}{\leq}  \norm{(\rho - \sigma)}_{\rm tr}\norm{O}_{\rm op}= 2\norm{O}_{\rm op} l(\rho,\sigma).
    \end{split}
\end{equation*}
The above inequalities are obtained by using the triangle inequality for the absolute value, and the H\"{o}lder inequality

\begin{equation*}
    \norm{AB}_{\rm 
    tr}= \norm{A}_{\rm p}\norm{B}_{\rm q}, \frac{1}{p}+\frac{1}{q}=1,
\end{equation*}
with $p=1$ and $q=\infty$, note that $\norm{X}_{\rm \infty}=\norm{X}_{\rm op}$ is maximal absolute value of all eigenvalues of X when X is Hermitian. (p = 1 correspond to the trace norm).

\section{OQSL for arbitrary dynamics}\label{app:E}

By using the Cauchy-Schwarz inequality $|{\rm tr}(AB)|$ $\leq$ $\sqrt{{\rm tr}(A^{\dagger}A){\rm tr}(B^{\dagger}B)}$ in Eq \eqref{OD}, we can obtain following inequality
 \begin{equation*}
  \cal{D}  \leq \frac{\sqrt{{\rm tr}(\rho ^2)}}{2\norm{O(0)}_{\rm op}}\sqrt{{\rm tr}[\{O(t) - O(0)\}^{\dagger}\{O(t) - O(0)\}]}.
\end{equation*}

The above inequality can be written in this form
\begin{equation}\label{ineq}
  \cal{D}  \leq \cal{D'} = \frac{\sqrt{{\rm tr}(\rho ^2)}}{2\norm{O(0)}_{\rm op}}\norm{\{O(t) - O(0)\}}_{\rm hs}.
\end{equation}

The rate of change of distance $\cal D'$ can be obtained by differentiating the above equation with respect to time. Thus, we obtain

\begin{equation*}
  \cal{\dot{D'}}  =\frac{\sqrt{{\rm tr}(\rho ^2)}}{2\norm{O(0)}_{\rm op}}\frac{{\rm tr}[\dot{O}(t)\{O(t) - O(0)\}+\{O(t) - O(0)\}\dot{O}(t)]}{2\norm{O(t) - O(0)}_{\rm hs}}.
\end{equation*}

The above inequality can be further simplified as
\begin{equation*}
  \cal{\dot{D'}}   = \frac{\sqrt{{\rm tr}(\rho ^2)}}{2\norm{O(0)}_{\rm op}}\frac{{\rm tr}[\dot{O}(t)\{O(t) - O(0)\}]}{\norm{O(t) - O(0)}_{\rm hs}}.
\end{equation*}

If we take the absolute value of $\cal{\dot{D'}} $ and again apply the Cauchy-Schwarz inequality $|{\rm tr}(AB)|$ $\leq$ $\sqrt{{\rm tr}(A^{\dagger}A){\rm tr}(B^{\dagger}B)}$, then we can obtain following inequality
\begin{equation*}
  |\cal{\dot{D'}} | \leq \frac{\sqrt{{\rm tr}(\rho ^2)}}{\hspace{2mm}2\norm{O}_{\rm op}}\norm{\dot{O}(t)}_{\rm hs}=\frac{\sqrt{{\rm tr}(\rho ^2)}}{\hspace{2mm}2\norm{O}_{\rm op}}\norm{L_{t}^{\dagger}({O}(t))}_{\rm hs}.
\end{equation*}

After integrating the above inequality, we obtain
\begin{equation*}
  \cal{{D'}}(O(T),O(0))  \leq \frac{\sqrt{{\rm tr}(\rho ^2)}}{\hspace{2mm}2\norm{O}_{\rm op}}T\Lambda_{T},
\end{equation*}
where $\Lambda_{T}=\frac{1}{T}\int_{0}^{T}dt{\norm{{L_{t}^{\dagger}({O}(t))}}_{\rm hs}}$ is evolution speed of observable $O$.

Let us use the Eq~\eqref{ineq}, then we obtain
\begin{equation*}
  \cal{{D}}(O(T),O(0))  \leq \frac{\sqrt{{\rm tr}(\rho ^2)}}{\hspace{2mm}2\norm{O}_{\rm op}}T\Lambda_{T}.
\end{equation*}

Finally, we obtain the desired bound on evolution time of expectation value of an observable as
\begin{equation}
T \geq  \frac{|\langle O(T) \rangle  -  \langle O(0) \rangle|}{\sqrt{{\rm tr}(\rho ^2)}\Lambda_{T}},
\end{equation}

where $T^{O}_{QSL}=\frac{|\langle O(T) \rangle  -  \langle O(0) \rangle|}{\sqrt{{\rm tr}(\rho ^2)}\Lambda_{T}}$.

This completes the proof of Eq \eqref{arbitrary}.

\section{OQSL for Dynamical Map}\label{app:F}
If the given quantum system has initial state $\rho$, and let its evolution is governed by a CPTP map ($\mathcal{E}$)
which is described by a set of kraus operators $\{K_{i}(t)\}$ and $\sum_{i}K^{\dagger}_{i}(t)K_{i}(t)=\mathcal{I}_{S}$. 
The dynamics of observable in the Heisenberg picture is described as 
\begin{equation}
O(t) =\sum_{i}K^{\dagger}_{i}(t)O(0)K_{i}(t).
 \end{equation}
The time evolution of expectation value of the observable $O$ is given by 
\begin{equation*}
\langle O(t) \rangle =\sum_{i} {\rm tr}[K^{\dagger}_{i}(t)O(0)K_{i}(t)\rho].
\end{equation*}

The rate of change of the expectation value of observable $O$ can be obtained by differentiating the above equation with respect to time, which is given by
\begin{equation*}
\begin{split}
\frac{ d\langle O(t) \rangle}{dt}
  = &\sum_{i}({\rm tr}[\dot{K}^{\dagger}_{i}(t)O(0){K}_{i}(t)\rho]\\ 
  &+  {\rm tr}[{K}^{\dagger}_{i}(t)O(0)\dot{K}_{i}\rho]).
  \end{split}
\end{equation*}

Let us take its absolute value, then we can apply the triangle inequality $|A+B|\leq|A|+|B|$ and the Cauchy-Schwarz inequality 
$|{\rm tr}(AB)|$ $\leq$ $\sqrt{{\rm tr}(A^{\dagger}A){\rm tr}(B^{\dagger}B)}$. Finally, we have obtained the following inequality 
\begin{equation*}
\begin{split}
  & \left|\frac{ d\langle O(t) \rangle}{dt}\right|\\
 &  \leq   \sum_{i}(\sqrt{{\rm tr}[\dot{K}^{\dagger}_{i}(t)O(0){K}_{i}(t)K^{\dagger}_{i}(t)O(0) \dot{K}_{i}(t)]{\rm tr}(\rho^2)}   \\
 &+ \sqrt{{\rm tr}[{K}^{\dagger}_{i}(t)O(0)\dot{K}_{i}(t)\dot{K}^{\dagger}_{i}(t)O(0){K}_{i}(t)]{\rm tr}(\rho^2)} ).
\end{split}
\end{equation*}

The simplified form of the above inequality is given as
\begin{equation*}
\begin{split}
   \left|\frac{ d\langle O(t) \rangle}{dt}\right| 
   \leq 2\sqrt{{\rm tr}(\rho^2)} \sum_{i}\norm{{{K}^{\dagger}_{i}(t)O(0)\dot{K}_{i}(t)}}_{\rm hs},
   \end{split}
\end{equation*}
where $\norm{A}_{\rm hs}$ = $\sqrt{{\rm tr}(A^{\dagger}A)}$ is the Hilbert-Schmidt norm of operator A.

After integrating the above inequality, we obtained the following bound
\begin{equation}
    T \geq   \frac{|\langle O(T) \rangle-\langle O(0) \rangle|}{2\sqrt{{\rm tr}(\rho^2)}\Lambda_{T}},
\end{equation}
where $\Lambda_{T}=\frac{1}{T}\int_{0}^{T}dt\norm{{K}^{\dagger}_{i}(t)O(0)\dot{K}_{i}(t)}_{\rm hs}$ is evolution speed of the observable $O$ and $T^{O}_{QSL}$= $\frac{|\langle O(T) \rangle-\langle O(0) \rangle|}{2\sqrt{{\rm tr}(\rho^2)}\Lambda_{T}}$ as given in Eq \eqref{Bound:open2}

For open dynamics, the bounds \eqref{Bound:open1}, and \eqref{Bound:open2} determines how fast the expectation value of an observable of the quantum system changes in time.
This completes the proof of Eq \eqref{Bound:open2}.

\section{QSL of two Point Function(Unitary Case)}\label{app:G}
Let us consider the two point correlation function, which is defined as
\begin{equation}
    C(t) = \langle A(t)A(0) \rangle - \langle A(t) \rangle \langle A(0) \rangle,
\end{equation}
where $A(t)$ = $U^{\dagger}(t)A(0)U(t)$ and $U^{\dagger}(t)U(t)$ = $I$.

After differentiating above equation with respect to time, we obtain 
\begin{equation}
   \frac{d C(t)}{dt} = \langle \dot{A}(t)A(0) \rangle - \langle \dot{A}(t) \rangle \langle A(0) \rangle. 
\end{equation}

Let us take the absolute value of the above equation and by using the triangular inequality $|A+B|\leq|A|+|B|$, we obtain
\begin{equation}
   \left|\frac{d C(t)}{dt}\right| \leq |\langle \dot{A}(t)A(0) \rangle| + |\langle \dot{A}(t) \rangle ||\langle A(0) \rangle|.
\end{equation}

Now, using the fact that $|{\rm tr}(A\rho)|$ $\leq$ $\norm{A}_{\rm op}$ (where $\rho$ is pure state), we can obtain the following inequality 
\begin{equation}
   \left|\frac{d C(t)}{dt}\right| \leq \norm{\dot{A}(t)A(0)}_{\rm op}  + \norm{ \dot{A}(t) }_{\rm op}\norm{ A(0)}_{\rm op}.
\end{equation}

The above equation can be expressed as
\begin{equation}
  \left|\frac{d C(t)}{dt}\right| \leq \norm{\dot{A}(t)}_{\rm op}\norm{A(0)}_{\rm op}  + \norm{ \dot{A}(t) }_{\rm op}\norm{ A(0)}_{\rm op}.
\end{equation}

This leads to
\begin{equation}
   \left|\frac{d C(t)}{dt}\right| \leq  2\norm{ A(0)}_{\rm op}\norm{ \dot{A}(t) }_{\rm op}.
\end{equation}

Therefore, we have
\begin{equation}
   \left|\frac{d C(t)}{dt}\right| \leq  \frac{2}{\hbar}\norm{ A(0)}_{\rm op}\norm{[H,A(t)] }_{\rm op}.
\end{equation}

After integrating, we obtain the following bound
\begin{equation}
 T \geq \frac{\hbar}{2} \frac{ |C(T) - C(0)|}{ \norm{ A(0)}_{\rm op} \frac{1}{T}\int_{0}^{T}dt\norm{[H,A(t)] }_{\rm op}}.
\end{equation}

This completes the proof of Eq \eqref{QSLTTU}.

\section{QSL of two Point Function (Open system Case)}\label{app:H}
Let us consider the two point function, which is defined as
\begin{equation}
    C(t) = \langle A(t)A(0) \rangle - \langle A(t) \rangle \langle A(0) \rangle,
\end{equation}

where  $A(t)=e^{\mathcal{L}^{\dagger}t}A(0)$ and ${\mathcal{L}^{\dagger}}$ is adjoint of Lindbladian.

After differentiating above equation with respect to time, we obtain 
\begin{equation}
   \frac{d C(t)}{dt} = \langle \dot{A}(t)A(0) \rangle - \langle \dot{A}(t) \rangle \langle A(0) \rangle.
\end{equation}
Lets take absolute value of the above equation and by using the triangular inequality $|A+B|\leq|A|+|B|$, we can obtain
\begin{equation}
   \left|\frac{d C(t)}{dt}\right| \leq |\langle \dot{A}(t)A(0) \rangle| + |\langle \dot{A}(t) \rangle ||\langle A(0) \rangle|.
\end{equation}

Now, using the fact that $|{\rm tr}(A\rho)|$ $\leq$ $\norm{A}_{\rm op}$ (where $\rho$ is pure state), we can obtain the following inequality 

\begin{equation}
   \left|\frac{d C(t)}{dt}\right| \leq \norm{\dot{A}(t)A(0)}_{\rm op}  + \norm{ \dot{A}(t) }_{\rm op}\norm{ A(0)}_{\rm op}.
\end{equation}

The above equation leads to
\begin{equation}
   \left|\frac{d C(t)}{dt}\right| \leq \norm{\dot{A}(t)}_{\rm op}\norm{A(0)}_{\rm op}  + \norm{ \dot{A}(t) }_{\rm op}\norm{ A(0)}_{\rm op}.
\end{equation}

The above equation can be written as
\begin{equation}
   \left|\frac{d C(t)}{dt}\right| \leq  2\norm{ A(0)}_{\rm op}\norm{ \dot{A}(t) }_{\rm op}.
\end{equation}

Therefore, we have
\begin{equation}
   \left|\frac{d C(t)}{dt}\right| \leq  \frac{2}{\hbar}\norm{ A(0)}_{\rm op}\norm{{\mathcal{L}^{\dagger}}[A(t)]}_{\rm op}.
\end{equation}

After integrating, we obtain the following bound
\begin{equation}
 T \geq \frac{\hbar}{2} \frac{ |C(T) - C(0)|}{ \norm{ A(0)}_{\rm op} \frac{1}{T}\int_{0}^{T}dt\norm{{\mathcal{L}^{\dagger}}[A(t)]}_{\rm op}}.
\end{equation}

This completes the proof of Eq \eqref{QSLTTO}.

\section{ QSL bound for commutators (Unitary Case)}\label{app:I}
The commutator of operators in two different regions of a many body system is defined as
\begin{equation}
    \langle{O(t)}\rangle = \langle[B(0), A(t)]\rangle,
\end{equation}
where $A(t)$ = $U^{\dagger}(t)A(0)U(t)$ and $U^{\dagger}(t)U(t)$ = $I$.

After differentiating the above equation with respect to time, we obtain 
\begin{equation}
   \frac{d \langle{O(t)}\rangle}{dt} = \langle[B(0), \dot{A}(t)]\rangle.
\end{equation}

Lets take absolute value of the above equation
\begin{equation}
   \left|\frac{d \langle{O(t)}\rangle}{dt}\right| = |\langle[B(0), \dot{A}(t)]\rangle|.
\end{equation}

Now, by using $|{\rm tr}(A\rho)|$ $\leq$ $\norm{A}_{\rm op}$ (where $\rho$ is pure state), we can obtain following inequality 
\begin{equation}
   \left|\frac{d \langle{O(t)}\rangle}{dt}\right| \leq \norm{[B(0), \dot{A}(t)]}_{\rm op}.
\end{equation}

Let us use the inequality $ \norm{[O_{1}, O_{2}]}_{\rm op}\leq 2  \norm{O_{1}}_{\rm op}  \norm{ O_{2}}_{\rm op}$, then we obtain
\begin{equation}
   \left|\frac{d \langle{O(t)}\rangle}{dt}\right| \leq 2\norm{B(0)}_{\rm op}\norm{ \dot{A}(t)}_{\rm op}.
\end{equation}

The above equation can be written as
\begin{equation}
   \left|\frac{d \langle{O(t)}\rangle}{dt}\right| \leq \frac{2}{\hbar}\norm{B(0)}_{\rm op}\norm{ [H,A(t)]}_{\rm op}.
\end{equation}

After integrating, we obtain the following bound
\begin{equation}
T \geq  \frac{2}{\hbar} \frac{ |\langle{O(T)}\rangle|}{ \norm{B(0)}_{\rm op}\frac{1}{T}\int_{0}^{T}dt\norm{[H,A(t)]}_{\rm op}}.
\end{equation}

This completes the proof of Eq \eqref{CQSLU}.

\section{ QSL bound for commutators (Open system Case)}\label{app:J}
The commutator of operators in two different regions of a many body system is defined as
\begin{equation}
    \langle{O(t)}\rangle = \langle[B(0), A(t)]\rangle,
\end{equation}
where  $A(t)=e^{\mathcal{L}^{\dagger}t}A(0)$ and ${\mathcal{L}^{\dagger}}$ is adjoint of Lindbladian.

After differentiating above equation with respect to time, we obtain 
\begin{equation}
   \frac{d \langle{O(t)}\rangle}{dt} = \langle[B(0) \dot{A}(t)]\rangle.
\end{equation}
Lets take absolute value of the above equation
\begin{equation}
   \left|\frac{d \langle{O(t)}\rangle}{dt}\right| = |\langle[B(0), \dot{A}(t)]\rangle|.
\end{equation}

Now, by using $|{\rm tr}(A\rho)|$ $\leq$ $\norm{A}_{\rm op}$ (where $\rho$ is pure state), we can obtain following inequality 
\begin{equation}
   \left|\frac{d \langle{O(t)}\rangle}{dt}\right| \leq \norm{[B(0), \dot{A}(t)]}_{\rm op}.
\end{equation}
Let us use the inequality $ \norm{[O_{1}, O_{2}]}_{\rm op}\leq 2  \norm{O_{1}}_{\rm op}  \norm{ O_{2}}_{\rm op}$, then we obtain
\begin{equation}
   \left|\frac{d \langle{O(t)}\rangle}{dt}\right| \leq 2\norm{B(0)}_{\rm op}\norm{ \dot{A}(t)}_{\rm op}.
\end{equation}
The above equation can be written as
\begin{equation}
   \left|\frac{d \langle{O(t)}\rangle}{dt}\right| \leq \frac{2}{\hbar}\norm{B(0)}_{\rm op} \norm{{\mathcal{L}^{\dagger}}[A(t)]}_{\rm op}.
\end{equation}
After integrating, we obtain the following bound
\begin{equation}
T \geq  \frac{2}{\hbar} \frac{ |\langle{O(T)}\rangle|}{ \norm{B(0)}_{\rm op}\frac{1}{T}\int_{0}^{T}dt\norm{{\mathcal{L}^{\dagger}}[A(t)]}_{\rm op}}.
\end{equation}

This completes the proof of Eq \eqref{CQSLO}.

 \end{document}